\documentclass[final,authoryear,1p,times]{elsarticle}

\usepackage{amsmath,amssymb,mathtools,mathrsfs,xspace,subfigure}
\usepackage{amsfonts}
\usepackage[usenames,dvipsnames]{xcolor}
\usepackage[]{graphicx}
\usepackage{url}
\usepackage{hyperref}
\usepackage{fullpage}
\usepackage{url}
\usepackage{float}
\usepackage{bm}
\usepackage{soul}
\usepackage{natbib}
\usepackage{ulem}
\renewcommand{\Xi}{\chi}
\newcommand{\vK}{\mathcal{V}}
\newcommand{\upd}{\mathrm{d}}
\newcommand{\lb}{\ell_b}
\newcommand{\tDelta}{\tilde{\Delta}}
\newcommand{\thetac}{\theta_{\rm conn}}
\newcommand{\thetam}{\theta_{\rm max}}
\newcommand{\Lhub}{\epsilon L}

\graphicspath{{./Figures/}}

\begin{document}

\begin{frontmatter}



\title{Localization of deformation in the central hub of \emph{hub-and-spoke} kirigami}

\author[a]{Jason Barckicke\fnref{label2}}
\author[a,b]{Lucie Domino\fnref{label2}}
\author[c]{Qun Zhang}
\author[a,c]{Mingchao Liu\corref{cor5}}
\author[a]{Dominic Vella\corref{cor5}}
\fntext[label2]{J.B. and L.D. contributed equally to this work.}
\cortext[cor5]{Corresponding authors: m.liu.2@bham.ac.uk (M.L.); dominic.vella@maths.ox.ac.uk (D.V.)}

\address[a]{Mathematical Institute, University of Oxford, Woodstock Rd, Oxford, OX2 6GG, UK}
\address[b]{Aix Marseille University
, CNRS, Institut
Universitaire des Syst\`{e}mes Thermiques et Industriels,
Marseille 13453, France}

\address[c]{Department of Mechanical Engineering, University of Birmingham, Birmingham, B15 2TT, UK}

\begin{abstract}
A recent approach to the design of flexible electronic devices consists of cutting a two-dimensional sheet to form a central hub connected to several tapered `spokes', resembling the hub-and-spoke of a bicycle wheel. When radially compressed, the resulting cut sheet buckles  out-of-plane forming a structure whose three-dimensional shape can be chosen by designing the tapering of the spokes. While the deformation of the spokes in  this `hub-and-spoke' kirigami are approximately cylindrical (i.e.~zero Gaussian curvature and hence small elastic strain), this is not the case in the central hub. The central hub is deformed radially because of continuity with the spokes but, because of its own circular symmetry, it must develop Gaussian curvature, and hence strain. In this article we quantify this strain, focussing in particular on its magnitude and its location. We find that the strain is localized in a boundary layer near the edge of the hub region, whose size is controlled by the moment applied on it by the deformed spokes. We discuss the implications of our results for avoiding material failure in flexible-electronic devices.
\end{abstract}


\end{frontmatter}

\section{Introduction}
Thin elastic objects occur at a range of scales from the capsids of viruses to the domes of cathedrals. Regardless of scale, the large aspect ratio of such structures ($length/thickness = L/t \gg 1$) means that their deformations are usually assumed to be isometric \citep{landau1970theory,vella2019buffering}: the energetic cost of changing lengths and areas is so much larger than that of bending that deformation should involve bending rather than stretching as $t/L \rightarrow 0$. 
This preference for isometry means that the allowed deformations of thin elastic objects are heavily constrained by Gauss’ \textit{Theorema Egregium} \citep{gauss1828disquisitiones}; loosely speaking, the Gaussian curvature of an elastic surface should remain constant under deformation. Deformations that would require changes of Gaussian curvature anywhere other than at isolated points or curves are therefore ‘\textit{Geometrically Incompatible}’ with the original object. 
The energetic expense of deforming a real object in a geometrically incompatible manner is what makes domes such effective engineering structures \citep{duffy2023lifting}, while also explaining how curving the edge of a slice of pizza prevents it from drooping under its own weight \citep{taffetani2019limitations}.

Recent research has focused on understanding and detailing non-trivial, but geometrically compatible, deformation modes for elastic sheets and shells in the limit of vanishing thickness. This has led to the development of a library of canonical deformation modes that can be used to understand more complex deformations; examples include the \textit{developable} cone (or ‘\textit{d}-cone’) \citep{cerda1998conical} and excess cone (or ‘\textit{e}-cone’) \citep{muller2008conical} for thin sheets, as well as `mirror buckling’ for a spherical shell \citep{pogorelov1988bendings,gomez2016shallow}. 
Such deformations satisfy the constraints of the \textit{Theorema Egregium} and so are expected to be good models for how real objects deform, even when $t/L$ is finite (but small). This programme has had remarkable success and has been used to understand deformations in settings as diverse as intestinal anastomotic surgery \citep{fleischer2022emergent} and the thermal motion observed in graphene \citep{xu2014unusual}.

At the same time, there is considerable interest in being able to generate curved shapes from intrinsically flat objects. One strategy is to prescribe localized regions of stretching  \cite[e.g.~via applied pressure, see][for example]{Siefert2019}, thereby removing any restrictions from Gauss' Theorem. Another strategy is to instead ‘subvert’ the restrictions imposed by Gauss' theorem by cutting away material (i.e., the art of kirigami); in this way the material appears to allow changes of distance at little cost \citep{zhang2015mechanically,lee2018two,choi2019programming,liu2020tapered,fan2020inverse,zhang2022shape,kansara2023inverse} --- a development that has been exploited to create curved structures in flexible electronics \citep{cheng2023programming}, soft robotics \citep{yang2023morphing}, metamaterials \citep{huang2024integration}, as well as tunable solar cells \citep{lamoureux2015dynamic}. 
These strategies allow changes to the ‘\textit{Apparent Gaussian Curvature}’ \citep{vella2019buffering,liu2020tapered}. However, as in other `ideal' deformations, kirigami localizes stress and strain to small regions of the material, which can lead to the possibility of plastic damage. This is not acceptable in applications like flexible electronics and solar cells.

\begin{figure}[h!]
  \centering
  \includegraphics[width=1.0\textwidth]{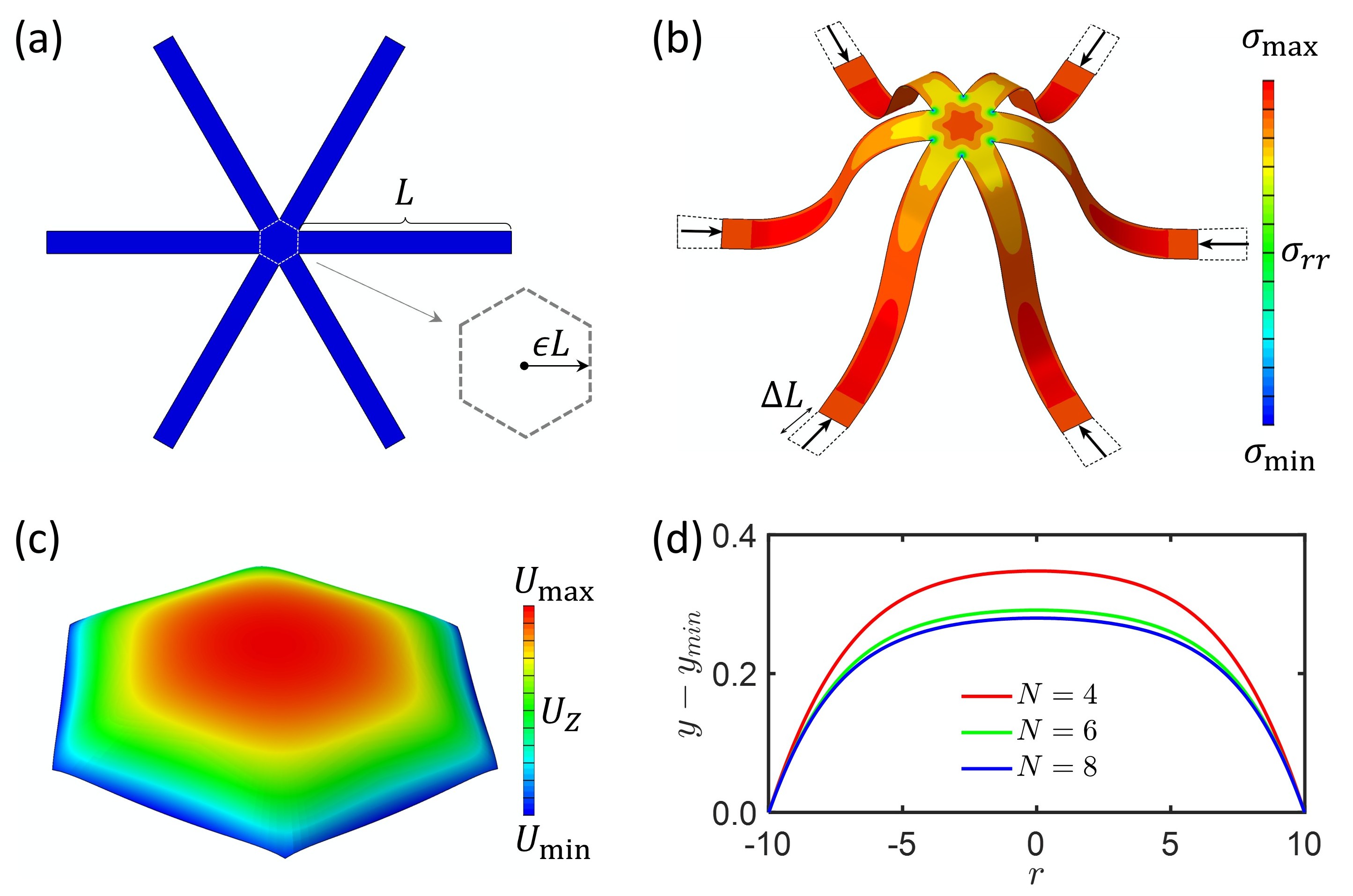}
  \caption{Schematic illustration of the \emph{hub-and-spoke} kirigami strategy for forming three-dimensional structures from a flat sheet and the corresponding deformed shapes.
  (a) The flat kirigami sheet with a central hub (radius $\epsilon L$) connected to six spokes (length $L$).
  (b) The deformed shape of the entire \emph{hub-and-spoke} kirigami subjected to radial end-shortening $\Delta L$ at the end of each spoke. The colormap indicates the radial stress in the axisymmetric coordinate system.
  (c) The deformed shape of the central hub. The colormap indicates the vertical displacement. It is displayed at ten times magnification for clearer visualization.
  (d) The cross-sectional shape of the central hub for kirigami with different numbers of spokes ($N = 4, 6$, and 8). Here, $L = 100$, $\epsilon = 0.1$, and $\Delta L = 40$ are fixed.
  }
  \label{fig:problemstatement}
\end{figure}

In this paper we focus on the hub-and-spoke type of kirigami  suggested by \cite{liu2020tapered} and \cite{fan2020inverse}, which is illustrated schematically in Fig.~\ref{fig:problemstatement}. Here, cuts are made to a two-dimensional (2D) sheet to form a shape with several `spokes' connected to a central `hub' region --- this cutting leads the sheet to resemble the hub-and-spoke arrangement of a bicycle wheel, as shown in Fig.~\ref{fig:problemstatement}a. The resulting shape buckles out-of-plane to form a three-dimensional (3D) structure when compressed in-plane (Fig.~\ref{fig:problemstatement}b); moreover, by choosing the tapering of the spokes, this buckled shape can be controlled \citep{liu2020tapered,fan2020inverse,zhang2022shape}.
In the deformed configuration, the hub-and-spoke kirigami shows very little stress in the legs, but considerable stress localization occurs where the spokes join the hub (see Fig.~\ref{fig:problemstatement}b), and is accompanied by the localization of the deformation around the edge of the hub region (Fig.~\ref{fig:problemstatement}c,d); the extent of this deformation, though not its localization, depends slightly on the number of `spokes' that are used (Fig.~\ref{fig:problemstatement}d). A similar focussing of stress at the join with a hub has also been reported in ribbon kirigami \cite[][]{Chen2024}. In this paper we seek to understand some features of this stress localization. To simplify the resulting analysis, we shall neglect any tapering of the spokes here and assume that the deformation of the central hub region is axisymmetric, as would be expected to be attained in the limit of a large number of spokes.

The paper is organized as follows. We start by introducing the problem statement in Section \ref{sec:ProblemStatement}. To describe the deformation of the whole structure, in Section \ref{sec:ModelEstablishment}, we present a model that combines the F\"{o}ppl-von K\'{a}rm\'{a}n equations (to describe the deformation in the hub region) with  the elastica equation (to describe the legs). 
The associated boundary conditions (BCs) to couple these two different models are discussed in some detail to close the problem and lead to the plate--elastica model. Numerical results for the plate--elastica model in the highly non-linear regime are presented in Section \ref{sec:HNL regime numerics} and compared to the results of finite element method (FEM) simulations for validation. The numerical results of the plate--elastica model suggest scaling laws for the angle at the connection between the hub and the legs, as well as the typical force scale. These numerical results also highlight that the hub region is very flat in the center with a narrow bent region around the edge. 
To further understand what makes the central hub so flat in the center, in Section \ref{sec:DecompositionMoment_Force}, we decompose the effect of the legs into an applied moment and an in-plane applied load. We show that the applied moment is more significant and present a boundary--layer solution; the scalings of this boundary layer solution explains  the observed scalings, as discussed in section \ref{sec:Scaling_spider-morph}.

\section{Problem statement}
\label{sec:ProblemStatement}

We consider perhaps the simplest type of hub-and-spoke kirigami in which each leg has length $L$ with uniform width $w$ and thickness $t$. \cite[In this way, we do not consider the leg tapering that was used to control the three-dimensional shape by][for example.]{liu2020tapered} Many such legs are connected to a  central `hub' region of the same thickness (see fig.\ref{fig:problemstatement}a). The radius of the central region is typically smaller than the length of each leg, and so we denote this radius by $\epsilon L$, where we expect that $\epsilon \ll 1$. In general, we expect that the legs may be inclined at non-small angles to the horizontal, $\theta=O(1)$, while the central hub region has much smaller inclinations to the horizontal, $\theta\ll1$. As such, it will be convenient to use the radial coordinate as the coordinate in the hub region, $0\leq r\leq \epsilon L$, while the position along the leg is best parametrized using  the arc-length, $\epsilon L\leq s \leq L(1+\epsilon) $. We consider a cross section through the kirigami pattern so that a general point has position $(x,y)$ and the local angle of inclination to the horizontal $\theta$.

We shall model the deflection of the legs using Euler's elastica \citep{Levien2008} with a horizontally clamped outer edge, $\theta((1+\epsilon)L)=0$, and an imposed horizontal load, $P$; we assume that $P$ is compressive, since this acts to actuate the hub-and-spoke kirigami pattern by inducing an inward radial displacement $\Delta L$. The central hub region is modelled as a circular plate (assuming that the number of legs $N\gg1$) --- we use the axisymmetric F\"{o}ppl-von K\'{a}rm\'{a}n (FvK) equations to describe the shape of this region. A key issue in this `plate-elastica model' is how to effectively get the elastica and FvK models to talk to each other at the boundary between the leg and the hub regions, $r=s=\epsilon L$; this will require careful consideration of the appropriate boundary conditions, informed by previous work on shell structures with piecewise constant natural curvatures \cite[see][for example]{liu2023snap}. We shall refer to the  edge between the central hub and the spoke as the ``connection" and will denote all quantities at the connection with a subscript ``conn'' (e.g.~$\thetac=\theta(\epsilon L)$). 

Our problem is to compute the shape of the deformed structure, particularly focussing on that in the central hub region where any deformation will, necessarily, be associated with an elastic strain and hence may have negative consequences in applications, for example in flexible electronic devices. The shape of the leg region is given by the leg's intrinsic equation, $\theta(s)$, while the deflection angle within the central region is $\theta(r)$. Within the leg region the stress is controlled by the imposed horizontal load, $P$, while in the hub region, the (thickness-integrated) stress state, $(\sigma_{rr},\sigma_{\phi\phi})$, is nonlinearly coupled to the hub's shape --- both must be determined together. 

\section{Development of the plate--elastica model}
\label{sec:ModelEstablishment}

\subsection{Equations for the hub and spoke deformations}

The origin of the coordinate system, $r=0$, is taken to be at the centre of the central hub. We assume that the hub is circular --- an approximation that is appropriate in the limit of a large number of spokes --- so that the edge of the hub is at  $r=\epsilon L$ (where $L$ is the length of the leg region). In this limit of a large number of spokes, the axisymmetric F\"{o}ppl-von K\'{a}rm\'{a}n (FvK) equations \cite[][]{Mansfield1989} for the deformation in the central region can be written in terms of the inclination angle, $\theta(r)$, and the derivative of the Airy stress function $\psi(r)$. (Here $\psi$ is defined such that the thickness-integrated radial and hoop stresses  $\sigma_r=\psi(r)/r$ and  $\sigma_{\phi}=\upd \psi/\upd r$, respectively.)

Integrating the vertical force balance equation (the first FvK equation) once and setting the constant of integration to zero (since there is no shear force at $r=0$) we have
\begin{equation}
    D\,r\frac{\upd}{\upd r}\left[ \frac{1}{r}\frac{\upd}{\upd r}\left(r\theta\right) \right]=\psi\,\theta,
    \label{eqn:FvK1Dim1}
\end{equation} where $D=Et^3/12\left(1-\nu^2\right)$ is the bending modulus of the sheet.  The first integral of the compatibility equation (the second FvK equation) gives
\begin{equation}
    r\frac{\upd}{\upd r}\left[ \frac{1}{r}\frac{\upd}{\upd r} \left(r\psi\right) \right]=-\frac{Y}{2}\theta^2,
    \label{eqn:FvK1Dim2}
\end{equation}
where $Y=Et$  is the stretching modulus of the sheet.

Our model of the central hub region assumes that the inclination angle remains small, $\theta\ll1$. While this is appropriate in this small central region, in general the angle $\theta=O(1)$ within the spoke region. For this reason, we use the \emph{elastica} to model deformations within each spoke, i.e.
\begin{equation}
\label{eqn:ElasticaDim}
    B\frac{\upd^2\theta}{\upd s^2}+P\sin\theta=0,
\end{equation}
where $B=Et^3/12$ is the bending modulus (per unit width) of the elastica, $P$ is the horizontal force (per unit width) at the edge, and $s$ is the arc-length coordinate. As the force is compressive, we have $P>0$. The choice of origin for the arc length coordinate is somewhat arbitrary (all that matters is that it should increase by $L$ along the length of the spoke); here we take $s\in[\epsilon L,(1+
\epsilon)L]$ since we envisage the central hub being subject only to small angle deflections for which the arc length is approximately equal to the radial coordinate. 

\subsection{Non-dimensionalization}\label{s:ndim equations}

Before analyzing the deformation or discussing the appropriate boundary conditions, we discuss how the governing equations \eqref{eqn:FvK1Dim1}--\eqref{eqn:ElasticaDim} are appropriately non-dimensionalized. A key feature is that we use different non-dimensionalizations in the central hub and spoke regions, which is important because it facilitates our study of the (small) central hub region.

\paragraph{Central Hub} In the central hub, we use the radius of the central hub, $\epsilon L$, as the relevant length scale while the relevant scale for the thickness-integrated stress is the typical buckling load per unit perimeter of the central hub, $D/(\epsilon L)^2$, so that $\psi$ is naturally scaled by $D/(\epsilon L)$ (since the stress involves derivatives of $\psi$).  We therefore introduce dimensionless variables
\begin{align}
\begin{aligned}
    R &= \frac{r}{\epsilon L} \in [0,1],\\
    \Psi &= \frac{\epsilon L}{D}\psi.
    \label{eq:BC_central}
\end{aligned}
\end{align} Upon non-dimensionalizating \eqref{eqn:FvK1Dim1}--\eqref{eqn:FvK1Dim2}, the dimensionless equations governing deformation in the hub region, $0\leq R\leq1$, are therefore
\begin{equation}
    R\frac{\upd}{\upd R}\left[ \frac{1}{R}\frac{\upd}{\upd R}(R\theta) \right]=\Psi\theta,
    \label{eq:Norm_plate_eq1}
\end{equation} and
\begin{equation}
    R\frac{\upd}{\upd R}\left[ \frac{1}{R}\frac{\upd}{\upd R}(R\Psi)\right]=-\vK\theta^2,
    \label{eq:Norm_plate_eq2}
\end{equation} where our non-dimensionalization has introduced a dimensionless parameter,
\begin{equation}
    \vK = \frac{(\epsilon L)^2}{2D}Y.
\end{equation} The parameter $\vK$ is proportional to the von K\'{a}rm\'{a}n number of the central hub \cite[see][for example]{NelsonSeung}; we shall use $\vK$ as a key parameter, and note that 
$\vK = 6\epsilon^2(1-\nu^2) L^2/t^2$ so that $\vK$ is a measure of the strip's slenderness.

\paragraph{Spoke region}

In the spoke region, we use the whole leg length $L$ to non-dimensionalize lengths and the typical buckling load, $P_*=B/L^2$, to non-dimensionalize tensions. We therefore introduce dimensionless variables
\begin{align}
\begin{aligned}
    S &= \frac{s}{L}+1-\epsilon \in [1,2],\\
    \mathcal{P} &= \frac{L^2}{B}P.
    \label{eq:BC_spoke}
\end{aligned}
\end{align}
(Note that in the definition of $S$, we have added a constant, $1-\epsilon$; this ensures that the outer edge of the central hub, $R=1$, corresponds to the inner edge of the elastica, $S=1$, and is included purely for convenience --- the choice of origin of $S$ is arbitrary.)

With this non-dimensionalization the elastica equation for the deformation of the spokes \eqref{eqn:ElasticaDim} becomes:
\begin{equation}
    \frac{\upd^2\theta}{\upd S^2}+\mathcal{P}\sin\theta=0
    \label{eq:Norm_elastica_eq}
\end{equation} for $1\leq S\leq 2$.

\subsection{Boundary conditions}\label{s:boundary conditions}

Clearly the trivial solution, $\theta=\Psi=0$, is an exact solution of the dimensionless system of equations  \eqref{eq:Norm_plate_eq1}, \eqref{eq:Norm_plate_eq2} and \eqref{eq:Norm_elastica_eq} for all values of $\mathcal{P}$. The system only buckles because a (dimensionless) compressive load, $\mathcal{P}$, is imposed at the far end of the spokes, i.e.~$S=2$. If $\mathcal{P}$ is sufficiently large it allows for out-of-plane deformation, $\theta\neq0$, and hence the end at $S=2$ moves a dimensional inward distance $\Delta L$; the relevant dimensionless measure of this distance is
\begin{equation}
\label{eqn:tDeltaDefn}
\tDelta=\Delta L/L=\int_1^2(1-\cos\theta)~\upd S-u(R=1)/L
\end{equation} where $u(R)$ is the radial displacement profile within the central hub. (In practice the contribution of $u(1)$ to $\tDelta$ is negligible in comparison to the integral.) 

For convenience, we shall treat the applied force $\mathcal{P}$ as a control parameter, together with the geometric parameters $\epsilon$ and $\vK$. We therefore have a sixth-order system (three second-order differential equations in the functions $\Psi$ and $\theta$) and so require six boundary conditions. We consider the three boundaries of the different region ($R=0$, $R=S=1$ and $S=2$) separately; we consider the connection between hub and spoke last.

\paragraph{The centre of the hub: $R=0$} Symmetry at the centre of the hub requires that
\begin{equation}
    \theta(R=0)=0,
    \label{eq:BC_axisymmetryCentral}
\end{equation}
 while the requirement of  zero radial displacement there leads to $0=u(R=0)=\lim_{R\to0}\left(R\epsilon_{\theta\theta}\right)$ and hence
\begin{equation}
\lim_{R\to0}    \left(R\frac{d\Psi}{dR}-\nu \Psi\right)=0,
\label{eq:BC_zerodisplacementCentral}
\end{equation} where $\nu$ is the Poisson ratio of the central hub.

\paragraph{The edge of the spoke: $S=2$} In most applications of the hub-and-spoke approach in, for example, flexible electronics, the edge of each spoke is adhered to a flat substrate. We shall therefore assume that the outer edge of each radial spoke is clamped, i.e.~that
\begin{equation}
    \theta(S=2)=0.
    \label{eq:BC_clampededge}
\end{equation}

\paragraph{The connection between hub and spoke: $R=S=1$} The remaining boundary conditions come from the connection  between the central hub and the spoke, which occurs at $R=S=1$.

A natural first condition is that the inclination angle must be continuous, i.e.
\begin{equation}
    \theta(R=1)=\theta(S=1):=\thetac,
    \label{eq:BC_continuityangle}
\end{equation}
where we have introduced $\thetac$ to denote the angle at the connection for later convenience.

In the same way, we also require continuity of the in-plane force,

i.e.~$\sigma_{rr}(r=\epsilon L)=-P\cos\bigl[\theta(s=\epsilon L)\bigr]$, which, when written in dimensionless notation gives
\begin{equation}
    \Psi(R=1)=-\epsilon^2\left(1-\nu^2\right)\mathcal{P}\cos\thetac.
    \label{eq:BC_continuityin-planeforce}
\end{equation}
Finally, the continuity of the bending moments, $M^+=M^-$ with $M^+=B\theta_s(s=\epsilon L)$ and $M^-=D\bigl[\theta_r(r=\epsilon L)+\nu\theta(r=\epsilon L)/(\epsilon L)\bigr]$, gives the dimensionless boundary condition
\begin{equation}
    \left. \frac{\upd\theta}{\upd R}\right|_{R=1}+\nu\theta(R=1) = 
    \left. \epsilon \left(1-\nu^2\right)\frac{\upd\theta}{\upd S}\right|_{S=1},
    \label{eq:BC_continuitymoment}
\end{equation} where we have assumed that the hub and spoke are made of the same material and have the same thickness, so that $B/D=1-\nu^2$.

We now have six boundary conditions \eqref{eq:BC_axisymmetryCentral}--\eqref{eq:BC_continuitymoment} for the sixth-order system comprising \eqref{eq:Norm_plate_eq1}, \eqref{eq:Norm_plate_eq2}, and \eqref{eq:Norm_elastica_eq}, making the system fully defined. However, it is reasonable to question whether the matching conditions imposed at the interface between the two distinct models (FvK and elastica) provide an accurate representation of physical reality. Therefore, before undertaking a more detailed asymptotic analysis of this model system, we turn to an alternative approach by validating the model numerically. Specifically, we compare its predictions with results from FEM simulations. This comparison will also highlight key features of the system's behavior, which we will later seek to explain analytically.

\section{Numerical validation of the model
\label{sec:HNL regime numerics}}

The relative end-end compression $\tDelta$, defined in \eqref{eqn:tDeltaDefn}, is intricately related to the in-plane compressive force imposed at the edge, $\mathcal{P}$ --- indeed, for the inextensible spokes considered here, a macroscopic value of $\tDelta$ is only possible for $\mathcal{P}>\mathcal{P}_c$, for some critical buckling load $\mathcal{P}_c$. In the plate-elastica model it is convenient to give $\mathcal{P}$ and compute the corresponding value of $\tDelta$. For $\mathcal{P}>\mathcal{P}_c$, we find that $\tDelta>\tDelta_{\min}$ and so in this section we consider only values $10^{-2}<\tDelta<0.7$, for which the structure is significantly compressed. Throughout this section  we shall take default values $\epsilon=0.1$, $\nu=0.3$,  and $t/L=10^{-3}$, which correspond to $\vK=5.46\times10^4$; we shall consider the effect of varying $\epsilon$.

\subsection{Shape of the structure}

To solve the system of equations \eqref{eq:Norm_plate_eq1}, \eqref{eq:Norm_plate_eq2} and \eqref{eq:Norm_elastica_eq} subject to the boundary conditions \eqref{eq:BC_zerodisplacementCentral}--\eqref{eq:BC_continuitymoment} we use the MATLAB routine \texttt{bvp4c}. We use continuation  on the parameter $\mathcal{P}$ starting from values for which a linearized solution of the elastica deformation is available. This allows the shape of the plate-elastica combination to be calculated for different parameter values. 

We begin by comparing the calculated out-of-plane deformation with that predicted from FEM simulations of the hub--spoke structures. The details of the FEM models are given in the Appendix. However, two details are important here. Firstly, the plate--elastica model is presented in terms of the inclination angle, $\theta$, while the FEM simulations present the shape $Y(X)$; for this comparison we therefore convert from $\theta$ to $X$ and $Y$ using
\begin{equation}
X=\begin{cases}R+u(R),\quad 0\leq R\leq 1\\
X(1)+\int_1^s\cos\theta~\upd S',\quad 1< S\leq 2,
\end{cases}
\end{equation}
and
\begin{equation}
Y=\begin{cases}Y(1)+\int^1_R\theta(R')~\upd R',\quad 0<R<1\\
\int_S^2\sin\theta~\upd S',\quad 1\leq S\leq 2.
\end{cases}
\end{equation} Note that we choose a sign convention such that $\upd Y/\upd S=-\sin\theta$ and $\upd Y/\upd R=-\theta$; this is more convenient since it avoids the necessity of discussing negative $\theta$ or reporting results for $|\theta|$.

Secondly, in FEM we must choose a number of spokes. If not otherwise stated, we take $N=6$ spokes in results shown here. We have also performed simulations with $N = 4$ and  $N=8$ spokes (see Fig.~\ref{fig:problemstatement}d, for example); these results indicate that as long as the  number of spokes is large enough (for example, $N >4$) the effect of $N$ on the deformed shape is negligible.

\begin{figure}[h!]
  \centering
  \includegraphics[width=1.0\textwidth]{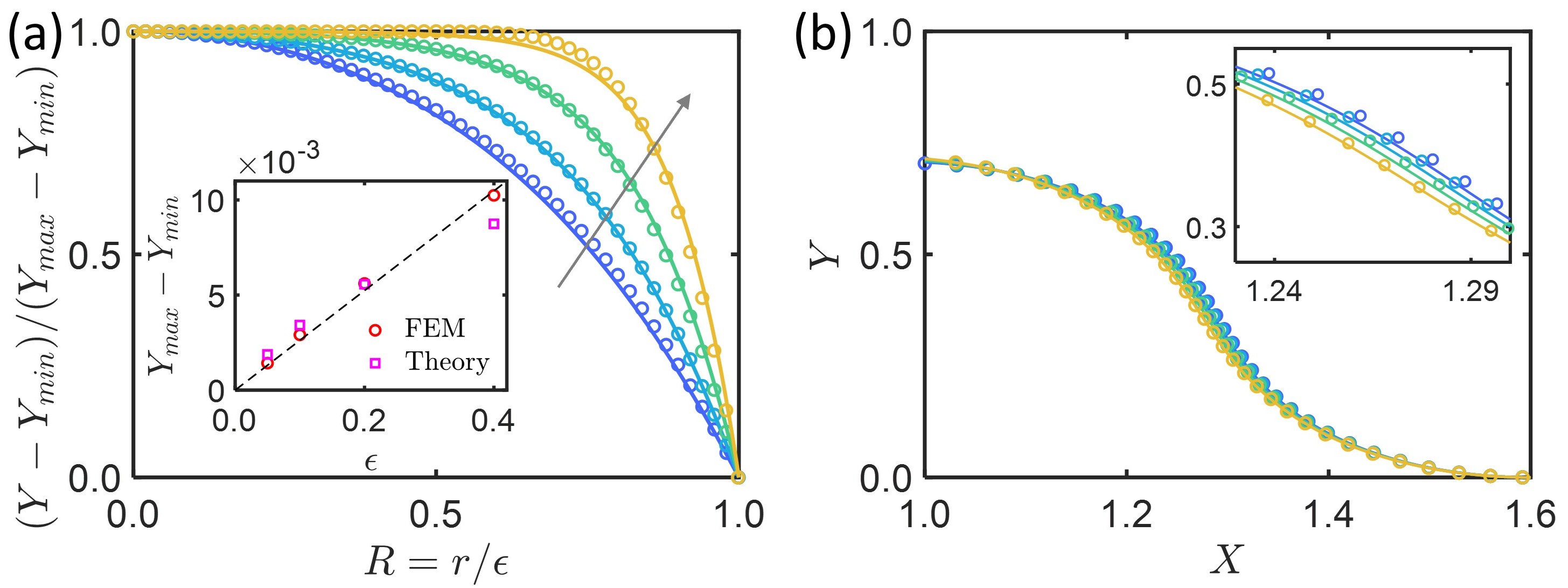}
  \caption{Comparison between the deformed shape of the \emph{hub-and-spoke} structure predicted by the FEM simulation (circles) and the plate--elastica model presented in \S\ref{sec:ModelEstablishment} (solid curves).
  (a) Normalized deformed shape of the central hub, $(Y - Y_{\min})/(Y_{\max} - Y_{\min})$, plotted as a function of the normalized radius $R = r/\epsilon$, for different values fo $\epsilon$. The arrow in the main figure shows the direction of increasing $\epsilon$; the inset shows $Y_{\max} - Y_{\min}$ as a function of $\epsilon$.
  (b) The deformed shape in the spoke region also shows good agreement between FEM simulations and the plate-elastica model. The inset shows a magnified view of the middle range for better comparison.
  In (a) and (b), $\nu=0.3$, $t/L=10^{-3}$, and  $\tDelta = 0.4$, with results 
  shown for $\epsilon = 0.05, 0.1, 0.2$, and $0.4$. (Hence $\vK=5.46\times10^4$ for $\epsilon=0.1$.)
  }
  \label{fig:deformedshape}
\end{figure}

The comparison between FEM and the plate--elastica model is shown in figure~\ref{fig:deformedshape} for a hub--spoke system with  a variety of values of the dimensionless hub size, $\epsilon$. Since the magnitude of the deformation in the hub region is much smaller than that in the spoke,  we separate the hub deformation (figure~\ref{fig:deformedshape}a) and the spoke deformation (figure~\ref{fig:deformedshape}b) to facilitate comparison between the FEM and model results. From figure~\ref{fig:deformedshape}a, we see that the agreement in the shape between the FEM and plate--elastica model is generally very good. In particular, we note that most of the curvature in the hub region is concentrated in a boundary layer near the connection to the spoke, i.e.~close to $R=1$, while the central hub is very flat in the center. Moreover, the width of this boundary layer becomes narrower for larger $\epsilon$. We also note that the vertical deflection at the centre, $Y_{\max}$, and at the connection, $Y_{\min}$, both vary over the four values of $\epsilon$ used, as shown in the inset of figure \ref{fig:deformedshape}a ,  but that the agreement between FEM and plate--elastica model remains good.
The deformed shapes of the spoke regions are presented in Figure \ref{fig:deformedshape}b. Again we see very good agreement between the FEM and plate--elastica model. In contrast with the hub region, however, the curves with different $\epsilon$ almost collapse on one another --- the size of the hub region does not significantly affect the deformation of the spoke region. 

Having seen that the two approaches agree well when describing the macroscopic shape of the deformed hub--spoke shape, we now turn to more detailed (and quantitative) comparisons with the numerical results. The angle at the connection between the hub and spoke region, $\thetac$, is of particular interest, since it gives a simple measure of how deformed the central hub region is. From a practical point of view, $\thetac$ also encodes the maximal strain within the hub region.

\subsection{The angle at the connection, $\thetac$}

A key measure of how deformed the central hub is by the buckling of the spokes is the angle at the connection, $\thetac$ --- since the origin of deformation in the hub is the deformed spokes, $\thetac$ is the maximum deformation within the hub. Similarly,  $\thetac$ provides the missing boundary condition for the spoke deformation --- if $\thetac$ were determined, the deformation of the spoke is effectively determined.  In this section, we therefore focus on numerical determination of the dependence of $\thetac$ on the imposed end-shortening, $\tDelta$, and the von K\'{a}rm\'{a}n number, $\vK$. We will see that both types of numerical results suggest various scaling relationships that we will then seek to understand in the remainder of the paper. 

\begin{figure}[h!]
  \centering  \includegraphics[width=1.0\textwidth]{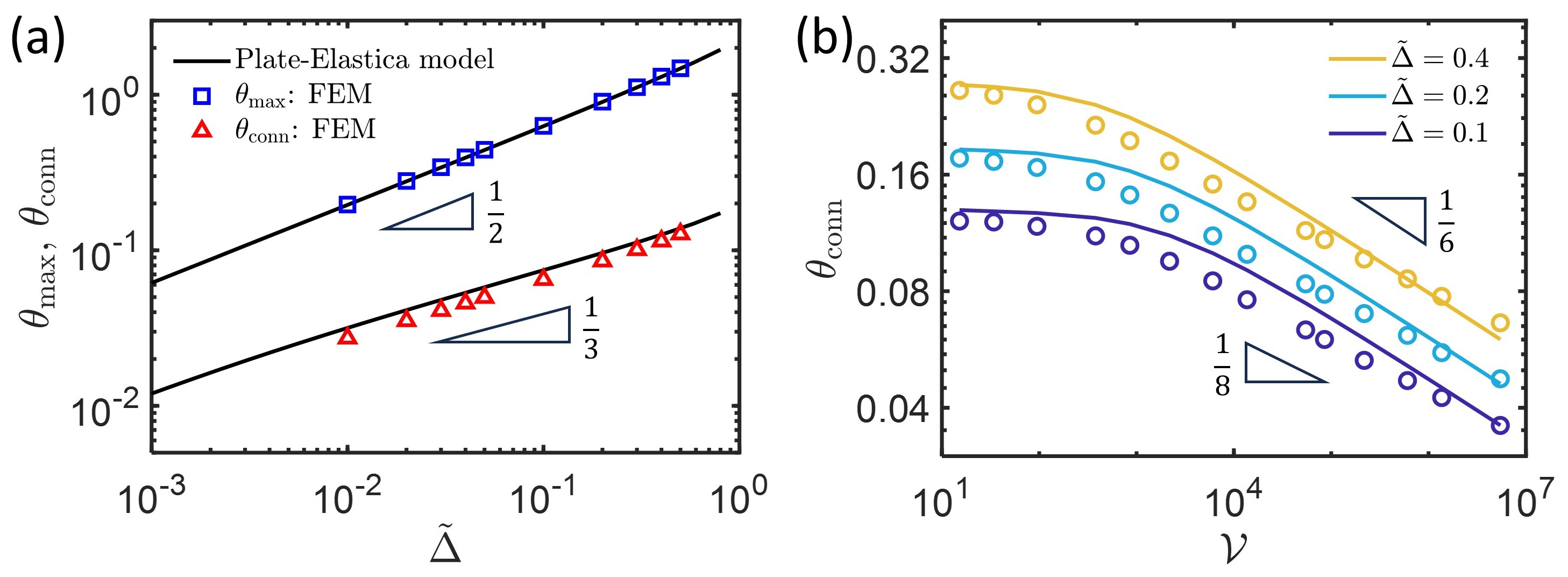}
  \caption{
  Comparison between FEM simulation results (symbols) and the numerical solution of the plate--elastica model (solid curves) for the angle at the connection between the spoke and hub, $\thetac$, and the maximum angle, $\thetam$, as a function of the imposed end-shortening $\tDelta$ and von K\'{a}rm\'{a}n number $\vK$. These numerical results suggest various scaling laws that motivate the rest of this paper. (a) The angle at the connection, $\thetac$, appears to exhibit a scaling with end-shortening $\tDelta$, namely $\thetac\propto\tDelta^{1/3}$, that is very different to the more standard relationship for the maximum angle $\thetam\propto\tDelta^{1/2}$ (also shown). Here, $t/L=10^{-3}$, $\epsilon=0.1$, $\nu=0.3$ so that $\vK=5.46\times10^4$ is fixed.
  (b) The angle at the connection $\thetac$ with fixed end-shortening $\tDelta\in\{0.1,0.2,0.4\}$ varies with the von K\'{a}rm\'{a}n number $\vK$. Here $\epsilon=0.1$, $\nu=0.3$ and $\vK$ is varied by changing $t/L$.
  }
  \label{fig:comparison theta edge}
\end{figure}

The angle at the connection, $\thetac$, is plotted as a function of the end-shortening, $\tDelta$, in figure \ref{fig:comparison theta edge}a. As for the deformed shape, this comparison shows good agreement between our theoretical results (solid curve) and the results of the FEM (the red triangles). We also note that both methods suggest  a range of values of $\tDelta$ in which $\thetac\propto\tDelta^{1/3}$. Interestingly, this scaling is distinct from the maximum angle in an elastica ---  a weakly nonlinear, or post-buckling, analysis \cite[see][for example]{Howell2009} predicts that $\theta_{\max}\propto \tDelta^{1/2}$, as is also observed in our numerical results (see blue squares in figure~\ref{fig:comparison theta edge}a). Note that in the regime we consider ($10^{-3} \ll \tDelta \ll 1$), $\thetac$ remains small, but $\theta_{\max}$ becomes of order unity. 

In addition to $\tDelta$, a key parameter determining the deformation of the central hub is the von K\'arm\'an number, $\vK$. Figure \ref{fig:comparison theta edge}b therefore also shows $\thetac$ for different $\vK$ at three different values of  $\tDelta$. As before, we find reasonable agreement between the predictions of the plate--elastica model (solid curves) and the FEM simulations (circles). Moreover, both approaches suggest a power-law trend for $\vK\gtrsim10^4$ in which $\thetac\propto \vK^{-\beta}$ and $1/8\lesssim \beta\lesssim 1/6$.

\section{Decomposition of the moment and the force experienced by the central hub}
\label{sec:DecompositionMoment_Force}

In the previous section we have shown that the plate--elastica model is able to accurately reproduce the results of more involved simulations using the FEM (ABAQUS). The key advantage of the plate--elastica model is that it allows us to analyse the governing equations and obtain a greater understanding of how the combined system behaves. A particularly striking feature of both numerical approaches is that the central hub is largely flat except for a narrow region close to the connection with the spoke in which  $\theta$ increases markedly. This localization of deformation also corresponds to a localization of strain (since the typical size of strain $\epsilon\sim \theta^2$) and so we now seek to understand \emph{why} the hub is so flat in general, but also how inclined its edge should be expected to be. 

To facilitate this understanding, we note that three of the six boundary conditions in \S \ref{s:boundary conditions} concern the connection $R=S=1$ and describe how the central hub responds to a horizontal force $\mathcal{P}$, a moment $\mathcal{M}$, and an imposed angle $\thetac$ from its interaction with the spoke region. While the imposition of the angle $\thetac$ at the boundary clearly leads to deformation, it is not immediately clear how the return of this deflection to zero is controlled: does the imposed moment or the imposed compression lead to the localization of deformation observed near the edge of the central hub? For this reason,  we separate the effects of a pure compressive force $\mathcal{P}$ and a pure moment $\mathcal{M}$ to better understand the deformations that they induce.

\subsection{Central hub under a pure compressive force $\mathcal{P}$ \label{sec:force}}

We consider the deformation of the central hub that is caused by a  compressive load per unit width, $\sigma_r=-\mathcal{P}\times B/(\epsilon L)^2$  but with zero imposed moment, i.e.~rather than \eqref{eq:BC_continuitymoment}, we impose
\begin{equation}
    \left.\frac{d\theta}{dR}\right|_{R=1} + \nu\theta_{R=1} = \mathcal{M} = 0.
    \label{eqn:zeroMoment}
\end{equation}

We therefore solve  the FvK equations, \eqref{eq:Norm_plate_eq1}--\eqref{eq:Norm_plate_eq2} with the boundary conditions \eqref{eq:BC_axisymmetryCentral}, \eqref{eq:BC_zerodisplacementCentral}, \eqref{eqn:zeroMoment} and the imposed load
\begin{equation}
    \Psi(R=1)=-\epsilon^2\left(1-\nu^2\right)\mathcal{P}
\end{equation} (assuming $\cos\thetac=1$ for small angles).

The resulting problem always admits the trivial solution $\theta(R)=0$ with $\Psi(R)=-\mathcal{P}\epsilon^2 \left(1-\nu^2 \right)R$. Above some critical load $\mathcal{P}_c$, however, a non-trivial solution exists. The linearized buckling analysis gives:

\begin{equation}
    \theta_{\rm buckle}(R)=A J_1(\Pi R)
\end{equation} where $\Pi=\epsilon(1-\nu^2)^{1/2}\mathcal{P}_c^{1/2}$ satisfies

\begin{equation}
\frac{\Pi}{2}\left[J_0(\Pi)-J_2(\Pi)\right]+\nu J_1(\Pi)=0
\end{equation} and $A$ is  a constant of integration that is undetermined in the linearized buckling analysis. We find that for $\nu=0.3$, the smallest non-trivial value of $\Pi\approx2.04885$; we  also find that for other $\nu$ the variation in this root is very small, varying by $\lesssim 10\%$ of this value for $0\leq\nu\leq0.5$.

\begin{figure}[h!]
  \centering
  \includegraphics[width=1.0\textwidth]{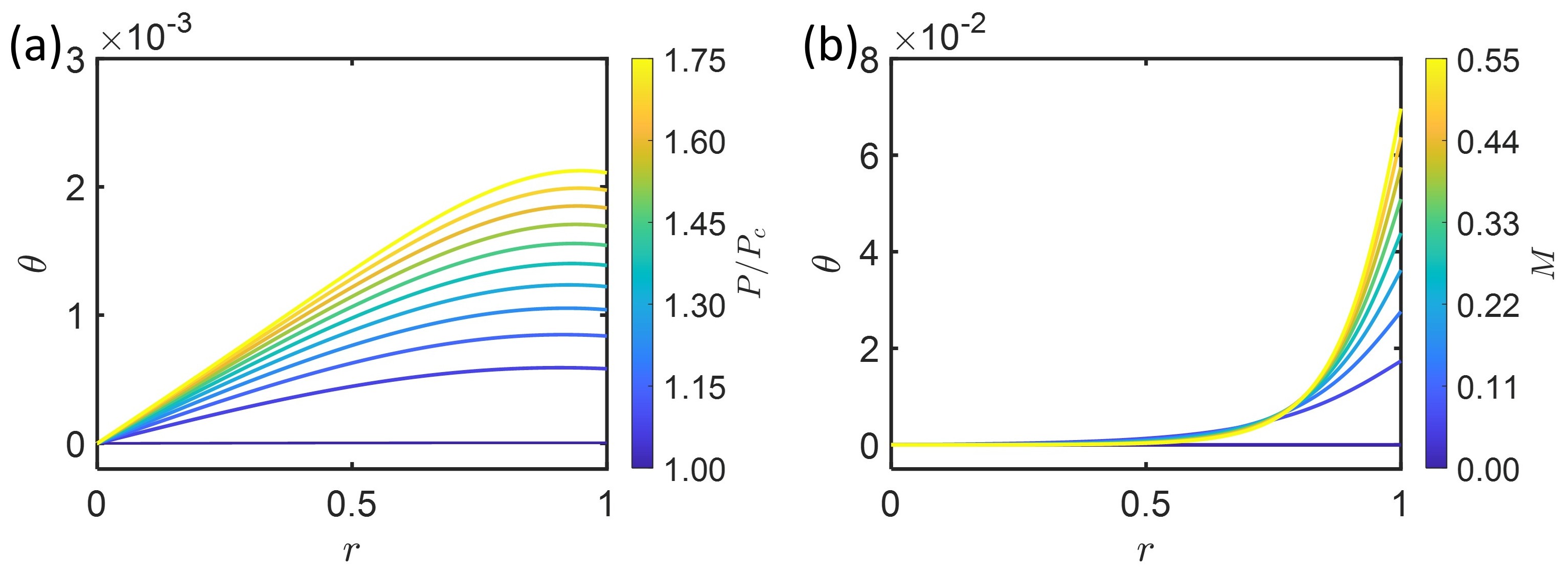}
  \caption{Results for the deformation of the central hub under the decoupled action of a compressive force and a bending moment at the outer edge. 
  (a) The distribution of $\theta_{\rm hub}(r)$ with given compressive force, $\mathcal{P}/\mathcal{P}_c \in [1,1.76]$ with linear spacing, $\mathcal{M}=0$ and $\vK=10^7$. 
  (b) The distribution of $\theta_{\rm hub}(r)$ with given bending moment, $\mathcal{M}\in [0,0.56]$ with linear spacing, $\mathcal{P}=0$ and $\vK=10^7$. 
  }
  \label{fig:Decomposition}
\end{figure}

Figure \ref{fig:Decomposition}a below shows the numerically computed shape for  $\theta(R)$ in the central hub with $\mathcal{M} = 0$ but a number of different values of the compressive force $\mathcal{P}$, with $1\leq{\mathcal{P}}/{\mathcal{P}_c}\leq1.76$. (The maximal value ${\mathcal{P}}/{\mathcal{P}_c} = 1.76$ has been chosen to be the value reached in the hub-and-spoke kirgami problem for $\tDelta = 0.7$; this will allow a direct comparison with the results obtained with a pure edge moment in \S\ref{sec:moment}.) In this range of values of $\mathcal{P}$, no localization of the deformation appears --- $\theta\propto r$ as $r\to0$ suggesting that the deformed shape of the hub region is parabolic. Crucially, the shape is deformed over the entire extent of the central hub. 

\subsection{Central hub under a pure moment $\mathcal{M}$ \label{sec:moment}}

\subsubsection{Problem statement} 
We now proceed in a similar manner for the case of an applied edge moment $M$, or its dimensionless equivalent, $\mathcal{M}=M\times \epsilon L/D$, but with no compressive load, $\mathcal{P}=0$. Again, the deformation of the central hub is governed by the FvK equations, \eqref{eq:Norm_plate_eq1} and \eqref{eq:Norm_plate_eq2}, but is  to be solved with the symmetry boundary conditions \eqref{eq:BC_axisymmetryCentral} and \eqref{eq:BC_zerodisplacementCentral} as well as the no in-plane load condition
\begin{equation}
    \Psi_{R=1} =0,
\end{equation} and the imposed moment condition
\begin{equation}
        \left.\frac{\upd\theta}{\upd R}\right|_{R=1} + \nu\,\theta(1) = \mathcal{M}.
        \label{eqn:BC_FixedM}
\end{equation}

This problem differs from the in-plane load case because there is no trivial solution --- the inhomogeneity in \eqref{eqn:BC_FixedM} ensures that there must always be some deformation. Figure  \ref{fig:Decomposition}b shows $\theta(R;\mathcal{M{}})$  for moments $\mathcal{M}$ in the range $0<\mathcal{M}\leq0.56$ maintaining $\mathcal{P}=0$. (Again,  the maximal value $\mathcal{M}=0.56$ has been chosen to be that reached in the hub-and-spoke kirigami problem with $\tDelta=0.7$, allowing a comparison with the results from \S\ref{sec:force}.) Interestingly, we see that the deformation in this case is qualitatively different to the compressive load case: rather than being deformed over the whole central hub region, deformation  is localized to a region close to $R=1$.

By comparing the results in Figure  \ref{fig:Decomposition}a and Figure  \ref{fig:Decomposition}b, it is clear that the effect of a moment from the spoke, $\mathcal{M}$, rather than the compressive force, $\mathcal{P}$, is more important  in explaining the  observed flattening near the centre of the hub in spoke--hub kirigami.
This observation motivates us to investigate the role of the moment $\mathcal{M}$, which will ultimately allow us to determine scaling laws for the deformation of the central hub.

\subsubsection{Analytical approach\label{s:M-scaling laws}}

 Our numerical results suggested that the key ingredient needed to understand the scaling laws observed numerically, particularly for $\thetac$ as a function of $\tDelta$, is the bending moment applied by the spoke on the edge of the central hub. We therefore seek to understand the effect of this bending moment on the central hub more fully. In particular, we focus on the case of relatively large moments where, presumably, the central hub bends only close to its edge; in this way we hope to exploit a narrow  boundary layer.
 
 \paragraph{Boundary layer scaling}
 
 Our calculation is inspired by analysis of other bending boundary layers that form near the edge of inverted shells \citep{Libai1998, liu2023snap}. We begin by seeking a scaling for the size of that boundary layer, $\lb$, as well as for the typical sizes of $\theta$ and $\Psi$. We zoom in on the narrow region close to the outer edge, $1-R\ll1$, and introduce a general rescaling
\begin{equation}
R = 1-R_* \xi,~\Psi = \Psi_* \,\Xi,~\theta = \theta_* \Theta, 
\end{equation} where we anticipate that $R_*\ll1$ (to allow us to zoom into the edge region of the central hub). This also means that spatial derivatives dominate division by $R$, namely: ${\upd}/{\upd R} = -R_*^{-1}(\upd/\upd\xi)\gg 1/R\approx1$, and the FvK equations become, at leading order in $R_*$,
\begin{equation}
    \frac{\upd^2\Theta}{\upd\xi^2} = \Psi_*R_*^2\,\Theta\,\Xi,
    \label{eqn:FvK1BL}
\end{equation}
\begin{equation}
    \frac{\upd^2\Xi}{\upd\xi^2}= -\vK\theta_*^2R_*^2\Psi_*^{-1}\,\Theta^2,
    \label{eqn:FvK2BL}
\end{equation} while the moment  boundary condition becomes
\begin{equation}
     \left.\frac{\upd\Theta}{\upd\xi}\right|_{\xi=0}=-\mathcal{M}R_*\theta_*^{-1}.
     \label{eqn:MomentBC}
\end{equation} The three unknowns ($\theta_*,R_*$ and $\Psi_*$) can be chosen to eliminate the dependence on $\vK$ and $\mathcal{M}$ in \eqref{eqn:FvK1BL}--\eqref{eqn:MomentBC}; we therefore choose
\begin{equation}
R_*=\mathcal{M}^{-1/3}\vK^{-1/6},\quad \theta_*=\mathcal{M}^{2/3}\vK^{-1/6},\quad \Psi_*=\mathcal{M}^{2/3}\vK^{1/3}.
\label{eqn:BLscaling}
\end{equation} With this scaling, the width of the region over which the central region is deformed is $\lb$ where
\begin{equation}
    \lb \sim R_*= \mathcal{M}^{-1/3}\vK^{-1/6}.
\end{equation}

\begin{figure}[H]           
    \includegraphics[width=0.49\textwidth]{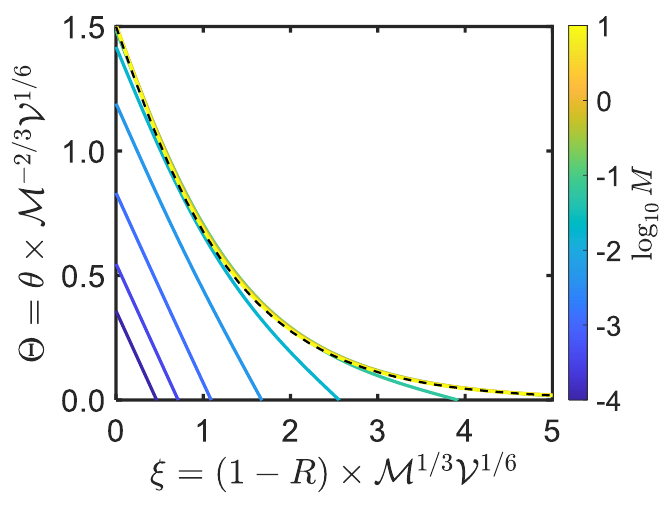}
    \centering
    \caption{Numerical results for $\Theta(\xi)$ in a central hub subject to a fixed edge moment $\mathcal{M}\in [10^{-4},10^1]$ and with $\vK=10^6$ fixed. Our boundary layer analysis predicts that as $\mathcal{M}\to\infty$, $\theta\sim \mathcal{M}^{2/3}\vK^{-1/6}$, while the deformation should be localized at the edge over a  horizontal length scale  $\lb\sim \mathcal{M}^{-1/3}\vK^{-1/6}$. Solid curves show numerical results of the edge-bending problem, rescaled according to \eqref{eqn:BLscaling}; the collapse of these curves with different imposed $\mathcal{M}$ confirm that the scalings of \eqref{eqn:BLscaling} are correct (compare with fig.~\ref{fig:Decomposition}(b) for example). The dashed black curve shows the  solution of the universal boundary layer-type problem, obtained by solving \eqref{eqn:FvK1BLFinal}--\eqref{eqn:BC_BL0} numerically. 
    }
    \label{fig:collapsed theta large M zoom}
\end{figure}

As a first check of these scalings, figure \ref{fig:collapsed theta large M zoom} shows numerical results with pure edge bending (e.g.~following the procedure used to plot figure \ref{fig:Decomposition}b but with different values of $\mathcal{M}$); these results are rescaled according to \eqref{eqn:BLscaling} and show very good collapse even with quite moderate values of $\mathcal{M}$.  This is aided by the factor $\vK^{-1/6}$, which ensures that the boundary layer width $\lb\sim R_*\ll1$ provided that $\mathcal{M}$ is moderately large.

\paragraph{Boundary layer problem}
We have now simplified the problem by rescaling $\mathcal{M}$ and $\vK$ out of the problem altogether; our rescaling used that $R_\ast\ll1$ at various points to determine a leading-order problem. In this limit, we then have
\begin{equation}
    \frac{\upd^2\Theta}{\upd\xi^2}=\Theta\,\Xi
    \label{eqn:FvK1BLFinal}
\end{equation}
\begin{equation}
    \frac{\upd^2\Xi}{\upd\xi^2}=-\Theta^2,
    \label{eqn:FvK2BLFinal}
\end{equation} which is to be solved subject to
\begin{equation}
    \Theta, \frac{\upd\Xi}{\upd\xi}\to0,\quad\mathrm{as}~\xi\to+\infty
    \label{eqn:BC_BLInfty}
\end{equation} (reflecting the boundary conditions at $R=0$) as well as
\begin{equation}
    \Xi|_{\xi=0}=0,\quad     \left.\frac{\upd\Theta}{\upd\xi} \right|_{\xi=0}=-1,
    \label{eqn:BC_BL0}
\end{equation} which correspond to the conditions at the edge that is subject to a constant moment, but no in-plane force.

\medskip
Unfortunately, we are not able to solve the system of equations \eqref{eqn:FvK1BLFinal}--\eqref{eqn:BC_BL0} analytically. Instead, we solve this system numerically (using the MATLAB routine \texttt{bvp4c}). The results compare very favourably with the numerical solution of the full problem, as shown in fig.~\ref{fig:collapsed theta large M zoom}. Moreover, we note that the numerical results tend to the universal solution provided that $(\mathcal{M}^2\vK)^{-1/6}\ll1$, which happens even with quite moderate values of $\mathcal{M}$ since $\vK$ is very large indeed. The most important result of this analysis is that
\begin{equation}
\label{eqn:thetac_W_moment}
    \thetac=\alpha\theta_*=\alpha \mathcal{M}^{2/3}\vK^{-1/6}
\end{equation} where the constant $\alpha\approx1.499$ is determined  from the numerical solution to \eqref{eqn:FvK1BLFinal}--\eqref{eqn:BC_BL0}.

We note that the scaling in \eqref{eqn:thetac_W_moment} recovers a $\thetac\propto\vK^{-1/6}$ power law that is consistent with our numerical results, from both the FEM and plate--elastica models --- see Fig.~\ref{fig:comparison theta edge}. However, at this stage the moment $\mathcal{M}$ is undetermined. We therefore turn to determine this moment by considering the coupling between the hub and the  spoke region. 

\section{Scaling laws in the plate--elastica model.}
\label{sec:Scaling_spider-morph}

Having  (a) shown that the flattening of the central hub region indicates a significant edge moment from the spoke/elastica on the central hub and (b) determined the scalings for the inclination of the edge in response to this moment, we now return to the full plate--elastica model. Our principal aim is to relate the moment imposed by the spoke, $\mathcal{M}$, to the imposed end-shortening of the spoke, $\tDelta$.

\subsection{Effective elastica and scaling predictions}

The calculation of the relationship between the connection angle $\thetac$ and the imposed moment $\mathcal{M}$ within the hub allows us to consider the elastica (spoke) region alone and, ultimately, to determine a relationship between $\mathcal{M}$ and $\tDelta$. To do this we consider the linearized spoke/elastica problem
\begin{equation}
    \frac{\upd^2\theta}{\upd X^2}=-\mathcal{P}\theta
    \label{eqn:LinElastica}
\end{equation} subject to boundary conditions
\begin{equation}
    \theta(1)=\thetac,\quad \theta'(1)=\frac{\mathcal{M}}{\epsilon(1-\nu^2)},\quad \theta(2)=0,
    \label{eqn:LinElastBC}
\end{equation} where we have used \eqref{eq:BC_continuitymoment} to obtain the boundary condition on $\theta'(1)$. The end-shortening constraint is then imposed by requiring that
\begin{equation}
    \tDelta=\int_1^21-\cos\theta~\upd X\approx\frac{1}{2}\int_1^2\theta^2~\upd X.
    \label{eqn:EndShorten}
\end{equation} Equation \eqref{eqn:EndShorten} will allow us to relate $\mathcal{M}$ to $\tDelta$, and hence determine the desired relationship between $\thetac$ and $\tDelta$.

We can immediately solve \eqref{eqn:LinElastica} subject to the boundary condition $\theta(2)=0$ to give
\[
\theta(X)=A\sin \bigl[\mathcal{P}^{1/2}(2-X)\bigr].
\] The boundary condition at $X=1$ gives
\[
\thetac=A\sin\mathcal{P}^{1/2}.
\] Given that both $A$ and $\mathcal{P}$ are unknown at this stage, we seem to have reached an impasse. However, we expect that $\thetac\ll A$ and so must have $\mathcal{P}^{1/2}\approx\pi$. It is possible to calculate the correction to this result, but this is not important for our purposes. Instead, we note that 
\[
\tDelta=\frac{A^2}{2}\int_1^2\sin^2\bigl[\mathcal{P}^{1/2}(2-X)\bigr]~\upd X=\frac{A^2}{4},
\] 
 so that $A\approx 2\tDelta^{1/2}$. This result immediately gives that the maximum angle within the elastica region, $\theta_{\max}\approx 2\tDelta^{1/2}$ --- a result that is indistinguishable from the numerical results presented in fig.~\ref{fig:comparison theta edge}a for $\tDelta\lesssim0.1$. Of more relevance here, however, is that we are now able to calculate the bending moment imposed on the hub by the end of the elastica, $\mathcal{M}=\theta'(1)\epsilon(1-\nu^2)=-A\epsilon(1-\nu^2)\mathcal{P}^{1/2}\cos\mathcal{P}^{1/2}\approx\pi\epsilon(1-\nu^2) A$; when combined with \eqref{eqn:thetac_W_moment}, this gives 
\[
\alpha^{-3/2}\thetac^{3/2}\vK^{1/4}=\mathcal{M}\approx\pi\epsilon(1-\nu^2) A\approx 2\pi\epsilon(1-\nu^2)\tDelta^{1/2},
\] and hence
\begin{equation}
\thetac\approx (2\pi)^{2/3}\alpha  \left[\epsilon(1-\nu^2)\right]^{2/3} \vK^{-1/6}\tDelta^{1/3}.    
\label{eqn:ThetaConnFinal}
\end{equation}

\subsection{Comparison of scaling predictions and numerical results}

\begin{figure}[H]
    \includegraphics[width=0.49\textwidth]{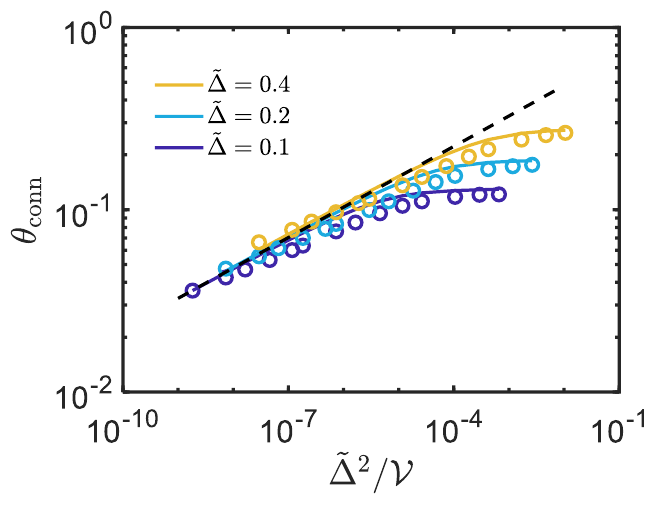}
    \centering
    \caption{ Comparison of the numerical solution of the plate–elastica model (solid curves) and the results of FEM simulations (points) with the scaling prediction of our analysis, \eqref{eqn:ThetaConnFinal}, (dashed line). Results are obtained for different values of $\tDelta$ and $\vK$ but all with $\epsilon=0.1$ and $\nu=0.3$.
    }
    \label{fig:Scaling laws in the spidermorph}
\end{figure}

 The relationship \eqref{eqn:ThetaConnFinal} gives us a closed form prediction for the angle at the connection between the central hub and the spoke, $\thetac$, in terms of both the imposed end--shortening, $\tDelta$, and the geometrical properties of the material (encapsulated through the von K\'{a}rm\'{a}n number $\vK$ and hub size $\epsilon$). A quick glance back at Fig.~\ref{fig:comparison theta edge} reveals that the scalings in \eqref{eqn:ThetaConnFinal} are consistent with the results of both our FEM simulations and the plate--elastica model. To demonstrate the quantitative agreement further, Fig.~\ref{fig:Scaling laws in the spidermorph} shows the results from Fig.~\ref{fig:comparison theta edge}b plotted as a function of $\tDelta^2/\vK$ ---\eqref{eqn:ThetaConnFinal} predicts that, with $\nu$ and $\epsilon$ held fixed, these results should collapse to a power law, $\thetac\propto (\tDelta^2/\vK)^{1/6}$. This collapse is reasonable provided that $\tDelta$ is small and $\vK$ is large. Moreover, the prefactor predicted by \eqref{eqn:ThetaConnFinal} is in good agreement with the data.

\subsection{Insight for flexible electronics}

Our study was motivated by the use of the hub-and-spoke kirigami strategy in morphing structures, particularly for flexible electronics. In this application, its purpose is to allow large deflection (generally of the spokes) without inducing significant strain. We have shown that this spoke deformation necessarily induces deformation (and hence strain) within the central hub region. We therefore now turn to consider the implications of our results for this original application.

We begin by noting that the central result of our analysis, \eqref{eqn:ThetaConnFinal} may be rewritten in terms of the underlying, dimensional, variables as:
\begin{equation}
\thetac\approx 3.79\, (1-\nu^2)^{1/2} \frac{\left(t\cdot \Delta L\cdot \Lhub\right)^{1/3}}{L}.    
\label{eqn:ThetaConnDim}
\end{equation} (Here, we have assumed that the material properties, such as the Poisson ratio $\nu$ and thickness $t$, of the spoke and hub are the same, for simplicity.)

The calculation of the inclination angle at the edge is key --- not only does it give a sense of the degree of the deformation of the central hub, but it also allows us to quantify the typical magnitude of the stretching strain $\epsilon_{\rm stretch}\propto\thetac^2$ \cite[see][for example]{Chopin2008}. This result  suggests the intuitively reasonable result that a good design principle to minimize the strains resulting from the hub-and-spoke kirigami concept is to keep the thickness and compression small, while making the spokes long. Perhaps less obviously, it also suggests that the hub region should be made as small as possible --- increasing $\Lhub$ allows the hub to bend more, which increases the strain that is localized at the hub edge (assuming all other variables are held fixed).

Another source of strain in this design is in the bending itself --- while the centerline of each spoke retains its initial length, the top and bottom surfaces are subject to a strain $\epsilon_{\rm bend}=t\cdot\theta'/2$. As a result, the maximum bending strain within the spoke is
\[
\epsilon_{\rm bend}\sim t\times \max_s\left(\frac{\upd \theta}{\upd s}\right)\approx 2\pi \frac{t}{L}\,\tDelta^{1/2}
\] and the ratio between the two is:
\[
\frac{\epsilon_{\rm bend}}{\epsilon_{\rm stretch}}\sim \left(\frac{t^2L^3}{(\Lhub)^4\,\Delta L}\right)^{1/6}.
\] As a result, we expect that the largest strain in the structure is observed in the spokes when 
\begin{equation}
    \tDelta\leq\tDelta_c\sim\frac{t^2L^2}{(\Lhub)^4},
    \label{eqn:tDeltaCrit}
\end{equation} but is expected to be in the hub region otherwise. Interestingly, this suggests that for small compression the strain is largest in the spoke region but then, as the spokes bend more, the effect that they have on the hub means that the strain in the hub region becomes dominant. Moreover, since $\tDelta<1$ (by definition), \eqref{eqn:tDeltaCrit} shows that if $\tDelta_c>1$ (i.e.~for sufficiently small hub regions, $\Lhub\lesssim (tL)^{1/2}$) the largest strain is always observed in the spokes, rather than the hub.

\section{Discussions and conclusions}

In this paper, we have considered the inevitable out-of-plane deformation that forms in the central hub region of the `hub-and-spoke' kirigami structures that have emerged in recent years \cite[][]{liu2020tapered,fan2020inverse,zhang2022shape,cheng2023programming,kansara2023inverse}. Since this  approach has generally been used to create 3D structures without significant stretching, quantifying the unavoidable out-of-plane deformation (and its associated strain), is important for understanding the limits on this approach.

We developed a mathematical model, the plate--elastica model, to describe how the central hub region interacts with the spokes in this geometry and validated this approach using FEM simulations of typical complete structures. Our model showed, unsurprisingly, that the effect of the hub region on the spoke's deformation is very limited, validating previous modeling work for the inverse problem that has assumed that the deformation of spokes is governed by Euler's elastica with clamped boundary conditions \cite[][]{liu2020tapered,fan2020inverse}. However, the forcing from the spoke region on the hub is significant. In particular, we showed that the non-zero bending moment exerted by the spoke on the edge of the hub is largely responsible for the deformation of the hub itself. Moreover, this deformation is localized to a typical distance
\begin{equation}
    \lb\sim\Lhub\cdot{\cal M}^{-1/3}\vK^{-1/6}\sim\left(\frac{L}{\Delta L}\right)^{1/6}t^{1/3}(\Lhub)^{2/3}
\end{equation} in the neighbourhood of the connection between hub and spoke. Together with this relationship, our main result is \eqref{eqn:ThetaConnFinal} --- or \eqref{eqn:ThetaConnDim} in dimensional form --- for the inclination angle at the connection between the hub and spoke regions (which is also the maximum inclination angle within the hub region).

The localized nature of the hub deformation has both positive and negative implications for the application of hub-and-spoke kirigami in morphing structures (particularly flexible electronic) applications. On the one hand, the localized nature of the deformation means that the hub is essentially flat over the majority of its area. On the other hand, the angle within the deformed region may become large and hence the associated strains may also be large enough to induce plastic deformation, or even material failure.

Finally, we note that the approach adopted here essentially patched together different elastic models (the plate and the elastica models) and then carefully studied the effective boundary conditions that act at the connection between the two. A similar approach has also had  success in determining how a spherical shell cap embedded within a plate deforms \cite[][]{liu2023snap}. As such this strategy seems likely to be more broadly applicable to understanding the stresses that form in kirigami patterns used in flexible and stretchable electronic applications \cite[see][for example]{lamoureux2015dynamic,kim2022}.

\subsection*{Acknowledgments}
The research leading to these results has received funding from the Royal Society through a Newton International Fellowship (M.L.), the start-up funding from the University of Birmingham (M.L.), and the Leverhulme Trust through a Philip Leverhulme Prize (D.V.). For the purpose of Open Access, the authors will apply a CC BY public copyright license
to any Author Accepted Manuscript version arising from this submission.


\end{document}